\documentclass[preprint,showpacs,preprintnumbers,amsmath,amssymb,
superscriptaddress]{revtex4}

\usepackage{graphicx}
\usepackage{dcolumn}
\usepackage{bm}

\begin{document}


\title{Search for $\beta^+$EC and ECEC processes in $^{112}$Sn}

\author{A.S.~Barabash} \email{barabash@itep.ru}
\affiliation{Institute of Theoretical and Experimental Physics, B.\
Cheremushkinskaya 25, 117218 Moscow, Russian Federation}
\author{Ph.~Hubert}
\affiliation{Centre d'Etudes Nucl\'eaires,
IN2P3-CNRS et Universit\'e de Bordeaux, 33170 Gradignan, France}
\author{A.~Nachab}
\affiliation{Centre d'Etudes Nucl\'eaires,
IN2P3-CNRS et Universit\'e de Bordeaux, 33170 Gradignan, France}
\author{S.I.~Konovalov}
\affiliation{Institute of Theoretical and Experimental Physics, B.\
Cheremushkinskaya 25, 117218 Moscow, Russian Federation}
\author{V.~Umatov}
\affiliation{Institute of Theoretical and Experimental Physics, B.\
Cheremushkinskaya 25, 117218 Moscow, Russian Federation}

\date{\today}

\begin{abstract}
Limits on $\beta^+$EC (here EC denotes electron capture) and ECEC processes 
in $^{112}$Sn have been obtained using 
a 380 cm$^3$ HPGe detector and an external source consisting of 53.355 g enriched
tin (94.32\% of $^{112}$Sn).
A limit with 90\% C.L. on the $^{112}$Sn half-life of 
$4.7\times 10^{20}$ y for the ECEC(0$\nu$) transition to the $0^+_3$ excited state 
in $^{112}$Cd (1871.0 keV) has been established. This transition is discussed 
in the context 
of a possible enhancement of the decay rate by several 
orders of magnitude given that the ECEC$(0\nu)$ process is nearly degenerate 
with an excited state in the daughter nuclide. Prospects for investigating 
such a process in future experiments are discussed.
The limits on other  $\beta^+$EC and ECEC processes in $^{112}$Sn  
were obtained on the level of $(0.6-8.7)\times 10^{20}$ y at the 90\% C.L.   
\end{abstract}



\maketitle

\section{Introduction}
Interest in neutrinoless $\beta\beta$ decay has seen a significant renewal in 
recent years after evidence for neutrino oscillations was obtained from the 
results of atmospheric, solar, reactor and accelerator  neutrino 
experiments (see, for example, the discussions in Refs. \cite{VAL06,BIL06,MOH06}). 
These results are impressive proof that neutrinos have a nonzero mass. However,
 the experiments studying neutrino oscillations are not sensitive to the nature
 of the neutrino mass (Dirac or Majorana) and provide no information on the 
absolute scale of the neutrino masses, because such experiments are sensitive 
only to the difference of the masses, $\Delta m^2$. The detection and study 
of $0\nu\beta\beta$ decay may clarify the following problems of neutrino 
physics (see discussions in Refs. \cite{PAS03,MOH05,PAS06}):
 (i) neutrino nature, 
whether the neutrino is a Dirac or a Majorana particle, (ii) absolute neutrino
 mass scale (a measurement or a limit on $m_1$), (iii) the type of neutrino 
mass hierarchy (normal, inverted, or quasidegenerate), and (iv) CP violation in 
the lepton sector (measurement of the Majorana CP-violating phases).
At the present time only limits on the level of $\sim 10^{24} - 10^{25}$ y 
for half-lives and $\sim 0.3-1$ eV for effective Majorana neutrino 
mass $\left<m_\nu\right>$ have been obtained in the best modern experiments 
(see recent reviews, Refs. \cite{BAR06,BAR08a,AVI07}).

The $\beta\beta$  decay can proceed through transitions  to the ground
state as  well as to various  excited states of  the daughter nucleus.
Studies of  the latter transitions provide supplementary
information about  $\beta\beta$ decay.  

Most $\beta\beta$ decay investigations have concentrated on the $\beta^-\beta^-$
 decay. Much less attention has been given to the investigation of 
$\beta^+\beta^+$, $\beta^+$EC, and ECEC processes (here EC denotes electron 
capture). There are 34 candidates for these processes. Only 6 nuclei can undergo
 all of the above-mentioned processes, 16 nuclei can undergo $\beta^+$EC 
and ECEC, and 12 nuclei can undergo only ECEC. Detection of the neutrinoless mode in
 the above processes enables one to determine the effective Majorana neutrino 
mass $\left<m_\nu\right>$ and parameters of right-handed current admixture in 
electroweak interaction ($\left<\lambda\right>$ and $\left<\eta\right>$). 
Detection of the two-neutrino mode in the above processes lets one determine 
the magnitude of the nuclear matrix elements involved, which is very important 
in view of the theoretical calculations for both the $2\nu$ and the $0\nu$ 
modes of $\beta\beta$ decay. Interestingly, it was demonstrated in Ref. 
\cite{HIR94} that if the $\beta^-\beta^-(0\nu)$ decay is detected, then the
 experimental limits on the $\beta^+EC(0\nu)$ half-lives can be used to obtain
 information about the relative importance of the Majorana neutrino mass and
 right-handed current admixtures in electroweak interactions.

The $\beta^+\beta^+$ and $\beta^+$EC processes are less favorable because of 
smaller kinetic energy available for the emitted particles and
Coulomb barrier for the positrons. However, an 
attractive feature of these processes from the experimental point of view is
 the possibility of detecting either the coincidence signals from four (two)
 annihilation $\gamma$ rays and two (one) positrons, or the annihilation 
$\gamma$ rays only. It is difficult to investigate the ECEC process because one
detects only the low energy x rays. It is also interesting to search for 
transitions to the excited states of daughter nuclei, which are easier to detect
 given the cascade of higher energy $\gamma$'s \cite{BAR94}. 
In Ref. \cite{WIN55} it was first mentioned that in the case of ECEC(0$\nu$)
transition a resonance condition can exist for transition to the "right energy"
of the excited level for the daughter nucleus; here the decay energy is close to zero.
In 1982 the same idea was proposed for the transition to the ground state 
\cite{VOL82}. In 1983 this possibility was discussed for the transition of
$^{112}$Sn to $^{112}$Cd ($0^+$; 1871 keV) \cite{BER83}. In 2004 the idea 
was reanalyzed in Ref. \cite{SUJ04} and new resonance conditions for the 
decay were formulated. 
The possible enhancement of the transition rate was estimated as $\sim$ 10$^6$
 \cite{BER83,SUJ04}. This means that this process starts to be competitive with
 $0\nu\beta\beta$ decay for the neutrino mass sensitivity and is  
interesting to check experimentally. There are several candidates for 
which resonance transition, to the ground ($^{152}$Gd, $^{164}$Eu and $^{180}$W)
 and to the excited states ($^{74}$Se, $^{78}$Kr, $^{96}$Ru, $^{106}$Cd, 
$^{112}$Sn, $^{130}$Ba, $^{136}$Ce and $^{162}$Er) of daughter nuclei, exists 
\cite{SUJ04,BAR07}. The precision needed to realize this resonance condition is well
 below 1 keV. To select the best candidate from the above list one must 
know the atomic mass difference with an accuracy better than 1 keV and
such measurements are planned for the future. Recently the experimental search
 for such a resonance transition in $^{74}$Se to $^{74}$Ge ($2^+$; 1206.9 keV) was
 performed yielding a limit of $T_{1/2} > 5.5\times10^{18}$ yr \cite{BAR07a}. Very
 recently $^{112}$Sn was investigated \cite{BAR08,DAW08,KID08}. The more strong 
limit of $T_{1/2} > 0.92\times10^{20}$ yr was obtained for the transition
to the $0^+$ state at 1871 keV with the 4-kg natural tin sample\cite{BAR08}.

In this article the results of an experimental investigation of the $\beta^+$EC and ECEC 
processes in $^{112}$Sn using the enriched tin sample are presented. 

\begin{figure*}
\begin{center}
\includegraphics[width=12cm]{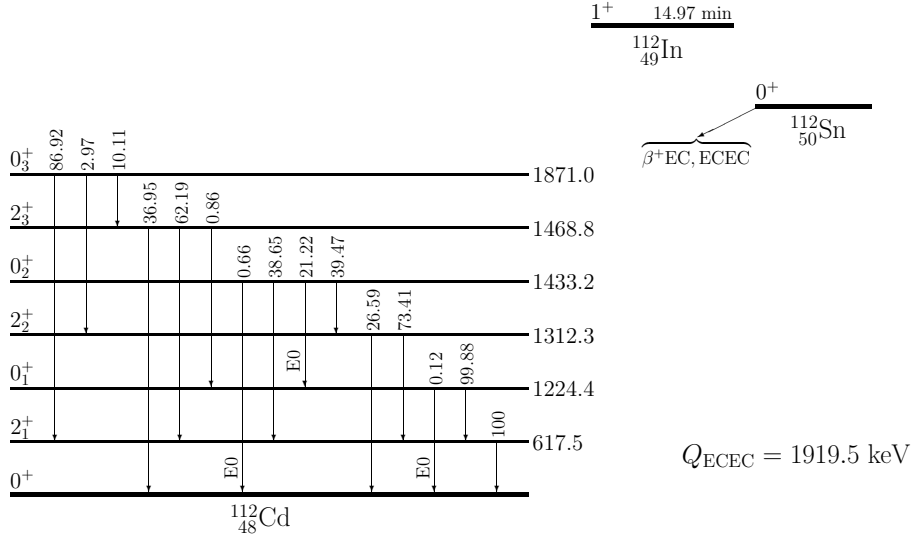}
\caption{\label{fig:fig1}Decay scheme of $^{112}$Sn. Only the investigated levels 
associated with $\gamma$ rays are shown. Transition probabilities are given in percentages.}  
\end{center}
\end{figure*}

\section{Experimental}

The experiment was performed in the Modane Underground Laboratory 
at a depth of 4800 m w.e. The enriched tin sample 
was measured using a 380 cm$^3$ low-background HPGe detector.

The HPGe spectrometer is a p-type crystal with  
the cryostat, endcap, and majority of mechanical components made of a very pure 
Al-Si alloy. The cryostat has a J-type geometry to shield the crystal from 
radioactive impurities in the dewar. The passive shielding consisted of 4 cm 
of Roman-era lead and 10 cm of OFHC copper inside 15 cm of ordinary lead. To 
remove $^{222}$Rn gas, one of the main sources of the background, a special 
effort was made to minimize the free space near the detector. In addition, 
the passive shielding was enclosed in an aluminum box flushed with radon-free 
air ($< 18$ mBq/m$^3$) delivered by a radon-free factory installed in the Modane 
Underground Laboratory \cite{NAH07}.

The electronics consisted of currently available spectrometric amplifiers and
an 8192 channel ADC. The energy calibration was adjusted to cover the energy 
range from 50 keV to 3.5 MeV, and the energy resolution was
2.0 keV for the 1332-keV line of $^{60}$Co. The electronics were stable during
the experiment because of the constant conditions in the laboratory (temperature 
of $\approx 23^\circ$ C, hygrometric degree of $\approx 50$\%).  A daily check
of the apparatus ensured that the counting rate was statistically constant.

The enriched tin sample, disk shaped (the diameter was 67 mm, the height was 2.2 mm),
was placed on the endcap of the HPGe detector. 
The sample mass was 53.355 g. Taking into account the enrichment of 94.32\%, in total
 50.3 g of $^{112}$Sn was exposed. 
 The duration of the measurement was 1885.8 h.

The sample was found to have a cosmogenic isotope, $^{113}$Sn ($T_{1/2}=115.09$ d),
with an average activity of $(18.8\pm 1.0)$ mBq/kg. The natural radioactivities had 
limits that were $<3.0$ mBq/kg of $^{226}$Ra,
$<4.6$ mBq/kg of $^{228}$Th, $< 27.2$ mBq/kg of $^{40}$K, and
$<1.2$ mBq/kg of $^{137}$Cs. 

The search for different $\beta^+$EC and ECEC processes in $^{112}$Sn 
were carried out using the germanium detector to look for $\gamma$-ray lines
 corresponding to these processes. 
The decay scheme for the triplet $^{112}$Sn-$^{112}$In-$^{112}$Cd is 
shown in Fig. \ref{fig:fig1} \cite{SIN96}. 
The $\Delta {\rm M}$ (difference of parent and daughter atomic masses) 
value of the transition is $1919.5. \pm 4.8$ keV \cite{AUD03}.
The following decay processes are possible:

\begin{equation}
e^-_b + (A,Z) \rightarrow (A,Z-2) + e^{+} + X   \hspace{2cm}  (\beta^+EC; 0\nu)   
\end{equation}

\begin{equation}
e^-_b + (A,Z) \rightarrow (A,Z-2) + e^{+} + 2\nu + X   \hspace{1cm}  (\beta^+EC; 2\nu)   
\end{equation}

\begin{equation}
2e^-_b + (A,Z) \rightarrow (A,Z-2) + 2X   \hspace{2.5cm}  (ECEC; 0\nu)   
\end{equation}

\begin{equation}
2e^-_b + (A,Z) \rightarrow (A,Z-2) + 2\nu + 2X   \hspace{1.5cm}  (ECEC; 2\nu)   
\end{equation}
where e$_b$ is an atomic electron and X represents x rays or Auger electrons.
Introduced here is the notation 
$Q'$ which is the effective $Q$ value defined as 
$Q'=\Delta {\rm M} - \epsilon_1-\epsilon_2$ 
for the ECEC transition and $Q'=\Delta {\rm M} - \epsilon_1- 2m_ec^2$ for the 
$\beta^+$EC process; $\epsilon_i$ is the electron binding energy of a daughter 
nuclide. For $^{112}$Cd, $\epsilon$ is equal to 26.7 keV for the K shell and 
4.01 keV, 3.72 keV and 3.54 keV for the L shell (2s, 2p$_{1/2}$ and 
2p$_{3/2}$ levels) \cite{BRO86}. In the case  of the L shell the resolution of the HPGe 
detector prohibits separation of the lines so we center the study on the 3.72 
keV line.

Investigations were made of the $\beta^+$EC transitions to the ground and the 2$^+_1$ 
excited states.
Additionally, the ECEC transitions to the ground state and six excited states 
(2$^+_1$, 0$^+_1$, 2$^+_2$, 0$^+_2$, 2$^+_3$ and 0$^+_3$) were investigated.

\begin{figure}
\begin{center}
\includegraphics[width=10.5cm]{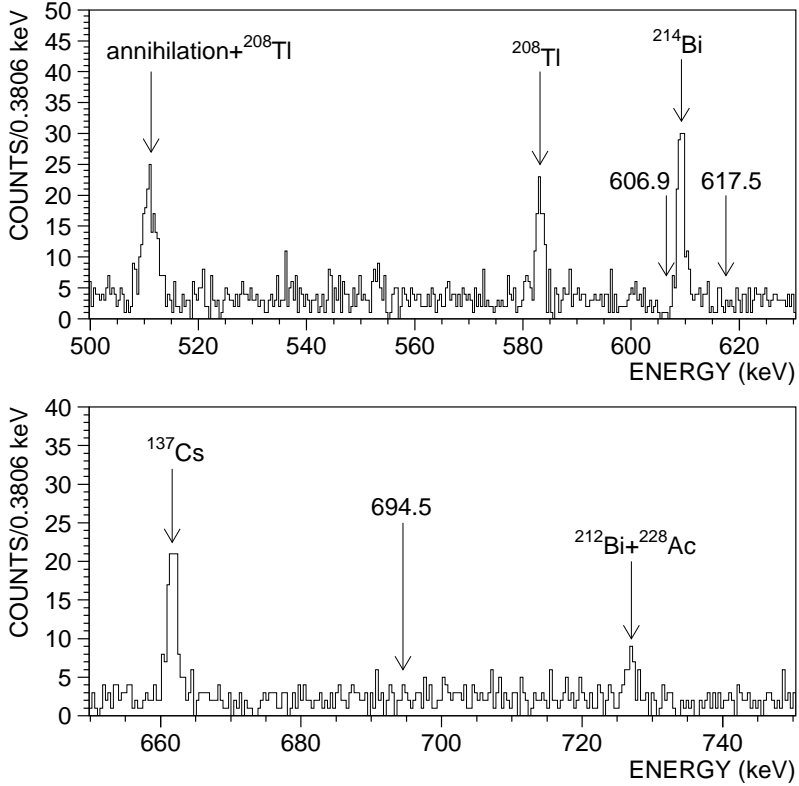}
\caption{\label{fig:fig2}Energy spectrum with 53.355 g of enriched Sn for 1885.8 h 
of measurement in the ranges investigated ([500-630] and [650-750] keV).}  
\end{center}
\end{figure}

\begin{figure}
\begin{center}
\includegraphics[width=10.5cm]{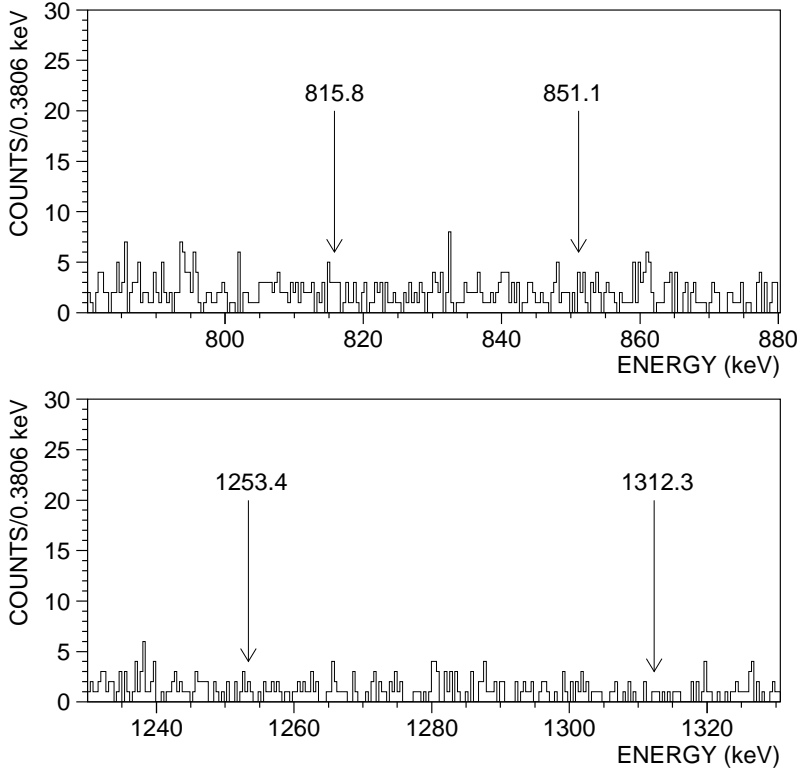}
\caption{\label{fig:fig3}Energy spectrum with 53.355 g of enriched Sn for 1885.8 h 
of measurement in the ranges investigated  ([780-880] and [1230-1330] keV).}  
\end{center}
\end{figure}

\begin{figure}
\begin{center}
\includegraphics[width=10.5cm]{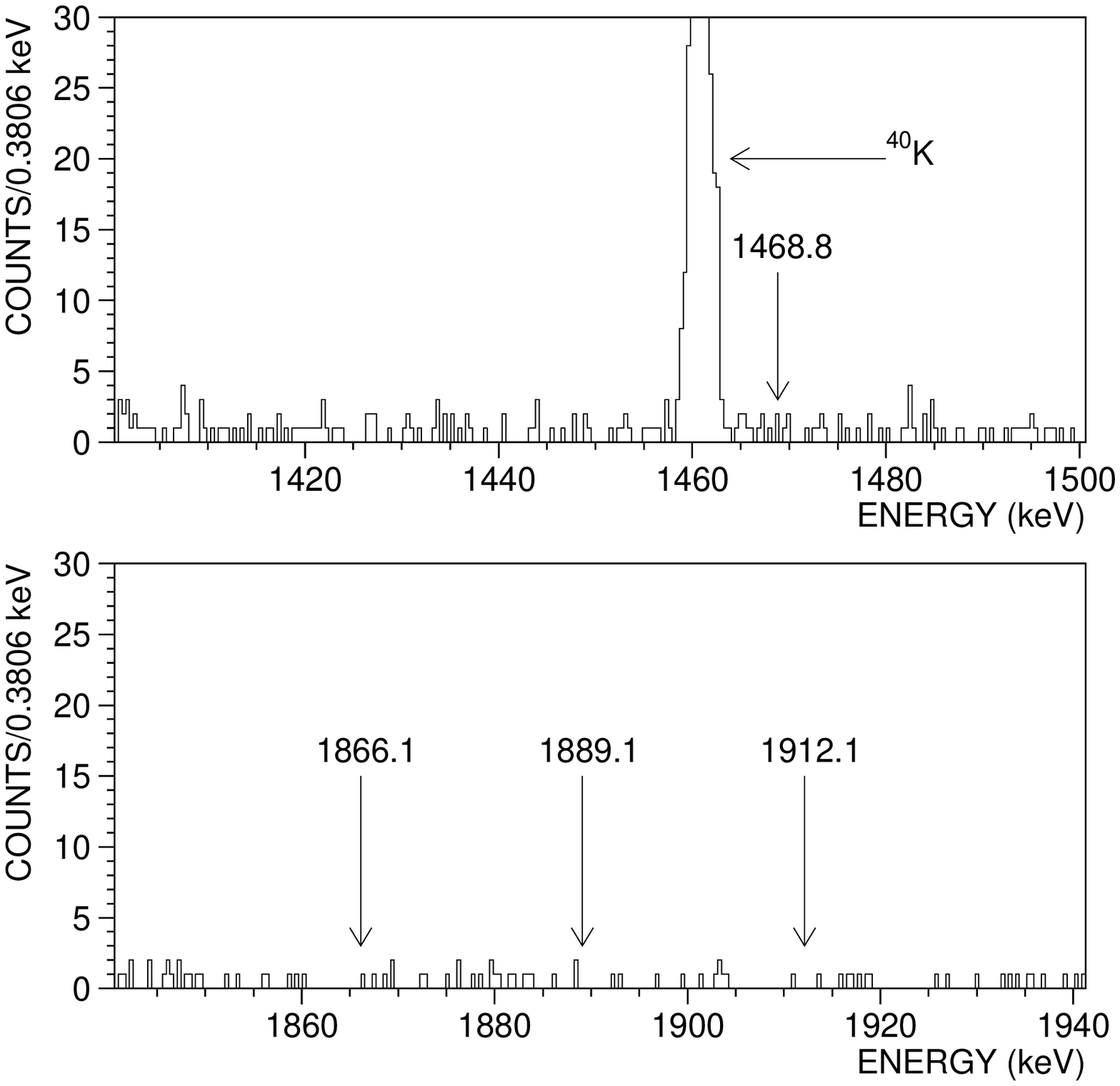}
\caption{\label{fig:fig4}Energy spectrum with 53.355 g of enriched Sn for 1885.8 h 
of measurement in the ranges investigated  ([1400-1500]  and [1840-1940] keV).}  
\end{center}
\end{figure}

The $\gamma$ ray spectra of selected energy ranges
 are shown in Figs. \ref{fig:fig2}, \ref{fig:fig3}, and \ref{fig:fig4}. 
These spectra correspond to regions of interest for 
the different decay modes of $^{112}$Sn. 

\subsection{ECEC transitions}  

The ECEC$(0\nu + 2\nu)$ transition to the excited states of $^{112}$Cd is 
accompanied with $\gamma$ quanta with different energies (see decay scheme in 
Fig. \ref{fig:fig1}). These $\gamma$ quanta were used in the search. 
The approach is not sensitive to ECEC$(2\nu)$ to the ground state
 because x rays are absorbed in the sample and cannot reach the sensitive 
volume of the HPGe detector.

The ECEC$(0\nu)$ transition to the ground state of the daughter nuclei was 
considered for three different electron capture cases:

1) Two electrons were captured from the L shell. In this case, $Q'$ was 
equal to $1912.1 \pm 4.8$ keV and the transition was accompanied by a 
bremsstrahlung $\gamma$ quantum with an energy of $\sim$ 1912.1 keV.

2) One electron was captured from the K shell and another from the L shell. 
In this case, $Q'$ was equal to $1889.1 \pm 4.8$ keV and the transition was 
accompanied by a bremsstrahlung $\gamma$ quantum with an energy of $\sim$ 
1889.1 keV.  

3) Two electrons were captured from the K shell. In this case, $Q'$ was equal
 to $1866.1 \pm 4.8$ keV and the transition was accompanied by a 
$\gamma$ quantum with an energy of $\sim$ 1866.1 keV. In fact this transition was 
strongly suppressed because of momentum conservation. So in this 
case the more probable outcome is the emission of an $e^+e^-$ pair \cite{DOI93}
that gives two annihilation $\gamma$ quanta with an energy of 511 keV.

The Bayesian approach \cite{PDG04} was used to estimate limits on transitions 
of $^{112}$Sn to the ground and excited states of $^{112}$Cd. To construct the 
likelihood function, every bin of the spectrum is assumed to have a Poisson 
distribution with its mean $\mu_i$ and the number of events equal to the 
content of the $i$th bin. The mean can be written in the general form,

\begin{equation}
\mu_i = N\sum_{m} {\varepsilon_m a_{mi}} + \sum_{k}
{P_k a_{ki}} + b_i .
\end{equation}

The first term in Eq. (5) describes the contribution of the investigated process 
that may have a few $\gamma$ lines contributing appreciably to the $i$th bin. 
The parameter $N$ is the number of decays, $\varepsilon_m$ is the detection 
efficiency of the $m$th $\gamma$ line and $a_{mi}$ is the 
contribution of the $m$th line to the $i$th bin. For low-background measurements a 
$\gamma$ line may be taken to have a Gaussian shape. The second term gives 
contributions of background $\gamma$ lines. Here $P_k$ is the area of the 
$k$th $\gamma$ line and $a_{ki}$ is its contribution to the $i$th bin. 
The third term represents the so-called ``continuous background'' 
($b_i$), which has been selected as a straight-line fit after rejecting all peaks 
in the region of interest. We have selected this region as the peak to be 
investigated $\pm$ 30 standard deviations ($\approx$ 20 keV). The likelihood 
function is the product of probabilities for selected bins.  
Normalizing over the parameter $N$ gives the probability density 
function for $N$, which is used to calculate limits for $N$.  To take into 
account errors in the $\gamma$-line shape parameters, peak areas, and other 
factors, one should multiply the likelihood function by the error probability 
distributions for these values and integrate to provide the average 
probability density function for $N$.

In the case of the ECEC$(0\nu)$ transition to the ground state of $^{112}$Cd 
there is a large uncertainty in the energy of the bremsstrahlung 
$\gamma$ quantum because of a poor accuracy in $\Delta M$ ($\pm$ 4.8 keV). 
Thus the position of the peak was varied in the region of the 
uncertainty and the most 
conservative value of the limit for the half-life was selected.       

The photon detection efficiency for each investigated process has been 
computed with the CERN 
Monte Carlo code GEANT 3.21. 
Special calibration measurements with radioactive sources and powders 
containing well-known 
$^{226}$Ra activities confirmed that the accuracy of these efficiencies is 
about 10\%.

The final results are presented in Table \ref{tab:tab1}. The fourth column shows the best 
previous experimental results 
from Ref. \cite{BAR08} for comparison. In the last column, 
the theoretical estimations for 
ECEC(2$\nu$) transitions obtained under the assumption of single intermediate 
nuclear state dominance 
are also presented \cite{DOM05}.

Concerning the ECEC$(0\nu)$ processes, the plan is to observe a resonant 
transition to the 1871.0 keV excited state of $^{112}$Cd. In this case we look 
for two peaks, at 617.5 and 1253.4 keV. In fact, 
the experimental spectrum has no extra events in the energy range of interest.
 The conservative approach gives the limit 
$T_{1/2} > 4.7\times 10^{20}$ yr at the 90\% C.L. 

\subsection{$\beta^+$EC transitions}

The $\beta^+$EC$(0\nu + 2\nu)$ transition to the ground state is accompanied 
by two annihilation $\gamma$ quanta with an energy of 511 keV. These 
$\gamma$ quanta were used to search for this transition. In the case of the 
$\beta^+$EC$(0\nu + 2\nu)$ transition to the 2$^+_1$ excited state the 
617.4 keV $\gamma$ quantum was also detected. To obtain limits on these 
transitions the analysis described in Sec. II A was used. Again the 
photon detection efficiencies for each investigated process was computed 
with the CERN Monte Carlo code GEANT 3.21. 
and are presented in Table \ref{tab:tab1}. 
The last two columns of the table show the best previous results and theoretical predictions
for comparison.
 
\begin{table*}
\caption{\label{tab:tab1}The experimental limits and theoretical predictions 
for the $\beta^+$EC and ECEC processes in 
$^{112}$Sn. $^{*)}$ For transition with irradiation 
of the $e^+e^-$ pair - see text.}
\begin{tabular*}{\textwidth}{lllll}
\hline

Transition & Energy of $\gamma$ rays, & \multicolumn{2}{c}{$T_{1/2}^{exp}$, 
$10^{20}$ y (C.L. 90\%)} & $T_{1/2}^{th}(2\nu)$, y \cite{DOM05}\\
\cline{3-4}
 & keV (Efficiency) & Present & Previous  \\
 &  & work & work \cite{BAR08} \\
\hline
$\beta^+$EC$(0\nu + 2\nu)$; g.s. & 511.0 (15.2 \%) & 0.56 & 0.12 & $3.8\times 10^{24}$ \\
$\beta^+$EC$(0\nu + 2\nu)$; 2$^+_1$ & 617.5 (3.92 \%) & 2.79 & 0.94 & $2.3\times 10^{32}$\\
\\
ECEC$(0\nu)$ ${\rm L}^1{\rm L}^2$; g.s. & 1912.1 (3.32 \%) & 4.10 & 1.3 \\
ECEC$(0\nu)$ ${\rm K}^1{\rm L}^2$; g.s. & 1889.1 (3.35 \%) & 3.55 & 1.8 \\
ECEC$(0\nu)$ ${\rm K}^1{\rm K}^2$; g.s. & 1866.1 (3.38 \%) & 3.97 & 1.3 \\
 & 511.0 (15.2 \%) & 0.59$^{*)}$ & 0.12$^{*)}$\\
\\
ECEC$(0\nu)$; 2$^+_1$ & 617.5 (5.53 \%) & 3.93 & 1.1 \\
ECEC$(0\nu)$; 0$^+_1$ & 606.9 (4.29 \%) & 6.87 & 1.2  \\
                      & 617.5 (4.25 \%) \\ 
ECEC$(0\nu)$; 2$^+_2$ & 617.5 (3.11 \%)  & 3.45 & 0.89 \\
                      & 694.9 (2.90 \%) \\
                      & 1312.3 (1.15 \%) \\          
ECEC$(0\nu)$; 0$^+_2$ & 617.5 (3.69 \%)  & 2.68 & 1.6   \\
                      & 694.9 (1.07 \%) \\
ECEC$(0\nu)$; 2$^+_3$ & 617.5 (2.64 \%)  & 2.64 & 0.93  \\
                      & 851.1 (2.09 \%) \\
                      & 1468.8 (1.34 \%) \\  
ECEC$(0\nu)$; 0$^+_3$ & 617.5 (5.09 \%)  & 4.66 & 0.92 \\
                      & 1253.4 (3.01 \%) \\
\\
ECEC$(2\nu)$; 2$^+_1$ & 617.5 (6.81 \%) & 4.84 & 1.2  & $4.9\times 10^{28}$\\
ECEC$(2\nu)$; 0$^+_1$ & 606.9 (5.42 \%) & 8.67 & 1.4  & $7.4\times 10^{24}$\\
                      & 617.5 (5.35 \%) \\

\hline
\end{tabular*}
\end{table*}

\addtocounter{table}{-1}
\begin{table}
\caption{Continued.}
\bigskip

\begin{tabular*}{\textwidth}{lllll}

\hline
Transition & Energy of $\gamma$ rays, & \multicolumn{2}{c}{$T_{1/2}^{exp}$, 
$10^{20}$ y (C.L. 90\%)} & $T_{1/2}^{th}(2\nu)$, y \cite{DOM05}\\
\cline{3-4}
 & keV (Efficiency) & Present & Previous \\
 &  & work & work \cite{BAR08}\\
\hline
ECEC$(2\nu)$; 2$^+_2$ & 617.5 (3.96 \%)  & 4.39 & 1.0  & $1.9\times 10^{32}$\\
                      & 694.9 (3.68 \%) \\
                      & 1312.3 (1.47 \%) \\        
ECEC$(2\nu)$; 0$^+_2$ & 617.5 (4.72 \%)  & 3.43 & 1.8  \\
                      & 694.9 (1.37 \%) \\     
ECEC$(2\nu)$; 2$^+_3$ & 617.5 (3.37 \%)  & 3.40 & 1.0  & $6.2\times 10^{31}$\\
                      & 851.1 (2.72 \%) \\
                      & 1468.8 (1.74 \%) \\  
ECEC$(2\nu)$; 0$^+_3$ & 617.5 (5.09 \%)  & 4.66 & 0.92 & $5.4\times 10^{34}$\\
                      & 1253.4 (3.01 \%) \\

\hline
\end{tabular*}
\end{table}

\section{Discussion} 

Limits obtained for the $\beta^+$EC and ECEC processes in $^{112}$Sn are on the level of 
$\sim (0.56-8.7)\times 10^{20}$ y or $\sim$ 2-5 times better 
than the best previous result \cite{BAR08} 
(see Table 1). 
As one can see from Table \ref{tab:tab1} the theoretical predictions 
for $2\nu$ transitions are much 
higher than the measured limits. The sensitivity of such experiments can still 
be increased with the 
experimental possibilities being the following:  
 \\ 1) Given 1 kg of enriched $^{112}$Sn in the setup described in   
Sec. II, the sensitivity after 1 yr of measurement will be $\sim 10^{22}$ yr. \\ 
2) With 200 kg of enriched $^{112}$Sn, using an installation such as GERDA \cite{ABT04} 
or MAJORANA \cite{MAJ03,AAL05} where 500-1000 kg of low-background HPGe detectors are 
planned, is a possibility. Placing $\sim$ 1 kg of very pure $^{112}$Sn around each of the 
$\sim$ 200 HPGe crystals both 
$^{76}$Ge and $^{112}$Sn will be investigated at the same time. 
The 
sensitivity after 10 yr of measurement may reach $\sim 10^{26}$ yr. Thus there is 
a chance of detecting the $\beta^+$EC(2$\nu$) transition of $^{112}$Sn to the ground 
state 
and the ECEC(2$\nu$) transition to the $0^+_1$ excited state (see theoretical predictions in 
Table I). 

In the case of the ECEC(0$\nu$) transition to the $0^+_3$ (1871.0 keV) excited state of 
$^{112}$Cd no extra events were detected. 
 So the search for this process continues into the future. 
Note that the ECEC(2$\nu$) transition to the $0^+_3$ excited state is strongly 
suppressed because of 
the very small phase space volume. In contrast, the probability of the 0$\nu$ 
transition should be 
strongly enhanced if the resonance condition is realized. In Refs. \cite{BER83,SUJ04} 
the "increasing factor" 
was estimated as $\sim 10^6$ and can be even higher. Then if the "positive" effect is 
observed in  
future experiments it is the ECEC(0$\nu$)
process. This will mean that lepton number is violated and the neutrino is a 
Majorana particle. To extract the
$\left<m_\nu\right>$ value one must know the nuclear matrix element for this 
transition and therefore the exact 
value of $\Delta {\rm M}$ (see Refs. \cite{BER83,SUJ04}). The necessary accuracy 
for $\Delta {\rm M}$ 
is better than 1 keV and this is  
a realistic task (in Ref. \cite{RED09} the $Q_{\beta\beta}({^{130}}$Te) was measured
 with an accuracy of 13 eV).

Two different descriptions for the resonance were discussed in the past. 
In Ref. \cite{BER83} the 
resonance condition is realized when $Q'$ is close to zero. They treat the process 
as (1S,1S) double 
electron capture and $Q'$ is equal to $-4.9 \pm 4.8$ keV (1$\sigma$ error). 
Thus there is a probability that $Q'$ is less than 1 keV. 
In this case one has a few 
daughter-nucleus $\gamma$ rays (see Fig.1) and two Cd K x rays, one of 
which may have 
its energy shifted by the mismatch in energies between the parent atom and the 
almost degenerate virtual 
daughter state. In Refs. \cite{SUJ04,LUK06}
the decay is treated as (1S,2P) double electron capture with irradiation of an 
internal bremsstrahlung photon. 
The $Q'$ value (energy of the bremsstrahlung photon) is $18.1 \pm4.8$ keV. 
The resonance condition for the transition 
is realized when $E_{brems}=Q_{res}=\mid E(1S,Z-2)-E(2P,Z-2)\mid $, i.e. when the 
bremsstrahlung photon energy becomes comparable to the $2P-1S$ atomic level 
difference in the final 
atom (23 keV). The same effect was theoretically predicted and then 
experimentally confirmed for 
single electron capture (see discussion in Ref. \cite{LUK06}).
It is anticipated, taking into account uncertainties in the $Q'$ value, that 
the real $Q'$ value is equal 
to 23 keV with an accuracy better then 1 keV and the resonance condition is realized. 
There are a few 
daughter-nucleus $\gamma$ rays (see scheme in Fig. \ref{fig:fig1}), one Cd K x ray and 
bremsstrahlung photon with energy 
$\sim K_{\alpha}$. 
The bremsstrahlung photon may have 
its energy shifted by the mismatch in energy between the parent atom and the 
almost degenerate virtual 
daughter state.

Finally, both approaches predict the same experimental signature for this 
transition and need to know with better accuracy 
the value of $\Delta {\rm M}$ to be sure that the resonance condition is really valid.
 New theoretical investigations of this transition are needed.

\section{Conclusion}

New limits on $\beta^+$EC and ECEC processes in $^{112}$Sn have been obtained using 
a 380 cm$^3$ HPGe detector and an external source consisting of 53.355 g enriched tin. 
In addition, it has been demonstrated that, in future larger-scale experiments, 
the sensitivity to the ECEC($0\nu$) 
processes for $^{112}$Sn can reach the order of $10^{26}$ yr. Under resonant 
conditions this decay will be competitive with $0\nu\beta\beta$ decay.

After submission of this article, we became aware of accurate $\Delta {\rm M}$ value measurements for  
$^{112}$Sn and $^{112}$Cd ($\Delta {\rm M}$ = $1919.82 \pm 0.16$ keV \cite{RAH09}). 
This result disfavors the strong enhancement scenario for the  
ECEC$(0\nu)$ process to the $0^+_3$ excited state 
in $^{112}$Cd.

\section*{Acknowledgements}
The authors thank the Modane Underground Laboratory staff for their 
technical 
assistance in running the experiment. 
Portions of this work were supported by a grant from RFBR 
(06-02-72553).
This work was supported by the Russian Federal Agency for Atomic Energy.

 --------------------------------------------------------------

\end{document}